\documentclass[american,aps,pra,reprint,longbibliography,superscriptaddress]{revtex4-2}

\usepackage{lmodern}
\usepackage[T1]{fontenc}
\usepackage[utf8]{inputenc}
\usepackage{color}
\usepackage{babel}
\usepackage{units}
\usepackage{amsmath}
\usepackage{amssymb}
\usepackage{graphicx}
\usepackage{hyperref}
 
\newcommand{\re}{\mathrm{e}}
\newcommand{\ri}{\mathrm{i}}
\newcommand{\rd}{\mathrm{d}}

\begin{document}

\title{Emergent topological properties in spatially modulated sub-wavelength barrier lattices}

\author{Giedrius \v{Z}labys}
\email{giedrius.zlabys@oist.jp}
\affiliation{Quantum Systems Unit, Okinawa Institute of Science and Technology Graduate University, 904-0495 Okinawa, Japan}
\author{Wen-Bin He}
\affiliation{Quantum Systems Unit, Okinawa Institute of Science and Technology Graduate University, 904-0495 Okinawa, Japan}
\author{Domantas Burba}
\affiliation{Institute of Theoretical Physics and Astronomy, Faculty of Physics, Vilnius University,  Saul\.{e}tekio 3, LT-10257 Vilnius, Lithuania.}
\author{Sarika Sasidharan Nair }
\affiliation{Quantum Systems Unit, Okinawa Institute of Science and Technology Graduate University, 904-0495 Okinawa, Japan}
\author{Thomas Busch}
\email{thomas.busch@oist.jp}
\affiliation{Quantum Systems Unit, Okinawa Institute of Science and Technology Graduate University, 904-0495 Okinawa, Japan}
\author{Tomoki Ozawa}
\affiliation{Advanced Institute for Materials Research (WPI-AIMR), Tohoku University, Sendai 980-8577, Japan}

\begin{abstract}
We investigate topological phenomena in a spatially modulated Dirac-$\delta$ lattice, where the scattering potential varies periodically in space. Changing the potential modulation frequency leads to Hofstadter's butterfly-like energy spectrum and enables the emergence of topological transport regimes characterized by non-trivial Chern numbers. We show how the considered modulated system is connected to the Hofstadter model via the Harper equation. By adiabatically varying spatial modulation parameters, we demonstrate controllable quantum transport and verify the topological nature of these effects through Wannier center displacement and bulk invariant calculations. We also propose an experimentally feasible realization of such a system using optically controlled three-level atoms. Our findings showcase spatially engineered Kronig-Penney-type systems as versatile platforms for investigating and exploiting different topological quantum transport regimes.
\end{abstract}

\maketitle

\section{\label{sec:Introduction}Introduction}

The Kronig-Penney model provides a paradigmatic description of crystalline solids with periodic potentials \cite{DeL1931Feb}. It is a one-dimensional continuum model in which a particle moves in a periodic array of short-range scatterers, often represented by a Dirac comb potential. 
It serves as an analytically tractable minimal description of low-dimensional condensed matter materials, capturing the physics of transport \cite{Sanchez1994Jan}, localization and disorder~\cite{Vaidya2023Sep,Lacki2023Nov,Izrailev2012Mar}. 
Recent theoretical and experimental developments in ultracold atomic systems allow such toy models to be realized in precisely controllable environments, with individually tunable positions and scattering strengths of the potential barriers.
Ultracold atomic systems offer a versatile platform where potentials with sub-wavelength structure can be engineered by using the spatial dependence of the nonlinear atomic response associated with the dark state of a three-level system~\cite{Lacki2016Nov,Wang2018Feb,Lacki2019Sep,Gvozdiovas2021Dec,Gvozdiovas2023Mar}, Fourier-synthesis of lattices utilizing multiphoton Raman transitions~\cite{Ritt2006Dec}, optical or radio-frequency dressing of optical potentials~\cite{Lundblad2008Apr} and trapping in near-field guided modes with nano-photonic systems~\cite{Gonzalez-Tudela2015May}.
These potentials are also supported by photonic crystals~\cite{Kuhl1998Apr,Mishra2003Jul}.

A central theme driving much of modern condensed matter physics is the profound role of topology.
The discovery of the quantum Hall effect revealed that quantum states could possess global properties, characterized by topological invariants, leading to robust phenomena such as quantized conductance immune to local perturbations~\cite{Klitzing1980Aug}.
Thouless pumping emerged as a powerful concept connecting the static topological properties with dynamic response~\cite{Thouless1983May}.
It demonstrated that the adiabatic and periodic change of system parameters could induce quantized transport, directly related to a topological invariant known as the Chern number.
This provided another experimental avenue for probing topology~\cite{Ozawa2019Mar, Citro2023Feb}.
Crucially, Thouless pumping allows to access high-dimensional systems by interpreting periodically varying parameters as synthetic dimensions.  
Thus, topological phenomena governed by Chern numbers can be engineered and studied even in systems that are effectively one-dimensional~\cite{Wang2013Jul, Nakajima2016Apr,Lohse2016Apr,Kraus2012Sep}. 

\begin{figure}[!t]
\includegraphics{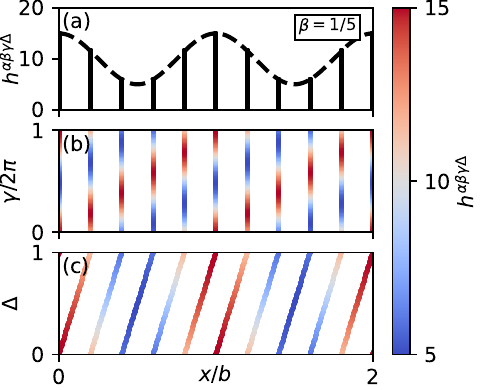}
\caption{\label{fig:fig0model} (a) Equidistantly spaced sub-wavelength barriers with scattering amplitudes $h^{\alpha \beta \gamma \Delta}$ for modulation parameters $\beta=\frac{1}{5}$, $\Delta = 0$, $\gamma=0$, $h_0=10$ and $\alpha=0.5$. The black dashed line indicates the modulation envelope. (b) Evolution of the barrier heights $h^{\alpha \beta \gamma \Delta}$ as $\gamma$ is changed while $\Delta = 0$. (c) Evolution of the barrier heights and positions as $\Delta$ is changed while $\gamma = 0$.
}
\end{figure}

Kronig-Penney-type models naturally lend themselves to such topological engineering. Parameters controlling modulation of the barrier positions or strengths can be mapped to synthetic dimensions, realizing effective higher-dimensional band structures with non-trivial topology. Moreover, these systems have been shown to support topological edge states~\cite{Reshodko2019Jan,Smith2019Nov,He2025Jan,Nair2025Mar}, demonstrating that non-trivial topology can be implemented and controlled at the level of continuum scattering potentials. The combination of analytical tractability, flexible tunability, and experimental accessibility therefore makes Kronig–Penney-type models an ideal setting to investigate topological transport.

In this work, we investigate the topological properties arising in a spatially modulated Dirac-$\delta$ lattice, where the scattering strengths of the equidistantly placed $\delta$-function potentials vary periodically in space. 
Specifically, we show that changing the frequency of this spatial modulation leads to the emergence of Hofstadter-like energy spectra, reminiscent of electrons in a two-dimensional lattice under a magnetic field.
We demonstrate that this spatial modulation coupled to the adiabatic variation of the modulation parameters enables control over quantum transport due to the presence of non-trivial topology in the system.
We explicitly calculate the Chern numbers of the spectrum, illustrating various possible transport regimes depending on the filling of the energy bands.
To solidify the connection to established topological frameworks, we compute the Wannier center displacement using different pumping protocols, confirming that they match the results obtained from the bulk topological invariant calculations and highlighting the role of the spatial barrier modulation as an effective magnetic flux that controls the periodicity length scale. 
Finally, we propose how to implement such a periodically modulated sub-wavelength barrier system in an experimentally accessible three-level dark state lattice.
Our work reveals how spatially structured potentials in simple lattice models can serve as a resource for engineering topological states and controlling quantum dynamics.

\section{\label{sec:Model}Model}
We consider a one-dimensional system of equidistantly spaced Dirac-$\delta$ scatterers (barriers) with spatially cosine modulated scattering amplitudes (heights). It is described by the dimensionless Hamiltonian 
\begin{equation}
\label{eq:model}
H = -\frac{\rd^2}{\rd x^2} +   \sum_{j\in \mathcal{M}} h^{\alpha \beta \gamma \Delta}_j   \delta(x-x_j^{\Delta}),
\end{equation}
where 
\begin{equation}
h^{\alpha \beta \gamma \Delta}_j \equiv h_0 [1+\alpha\cos(2\pi\beta x_j^{\Delta}-\gamma)].    
\end{equation}
Throughout the paper energy is taken to be dimensionless and is measured in terms of $E_0 = \frac{\hbar^2}{2m a^2}$, where $a$ is the separation between neighboring barriers, $m$ is the mass of the particle and $\hbar$ is the reduced Planck's constant. The position coordinate $x$ and the $j$-th barrier position $x_j^{\Delta}=x_j+\Delta$ are measured in units of $a$, where $x_j \equiv x_j^{0}$ and for numerical calculations we set $x_0 =0$. $\mathcal{M}\subset \mathbb{Z}$ is the indexing set of the barriers and the parameter $\Delta$ indicates a global translation of all of the scatterers by distance $\Delta$ under the height modulation function. The cosine modulation amplitude is $\alpha \in{[0,1]}$, and its spatial frequency and phase are $\beta$ and $\gamma \in [0,2\pi)$ (a shift of $\gamma = 2\pi\beta$ is equivalent to translating the whole lattice by scatterer separation distance $a$). When the modulation is turned off ($\alpha = 0$), the height of all barriers is $h_0$. An example spatial barrier configuration for $\beta=\frac{1}{5}$ is shown in Fig.~\ref{fig:fig0model}(a) for two elementary cells of length $b$. The evolution of the barrier positions and heights under periodic change of $\gamma$ and $\Delta$ are shown in Fig.~\ref{fig:fig0model}(b) and (c) respectively.

\section{\label{sec:Results}Results}

\subsection{\label{subsec:EnergySpectrum}Band structure}

\begin{figure*}[!ht]
\includegraphics{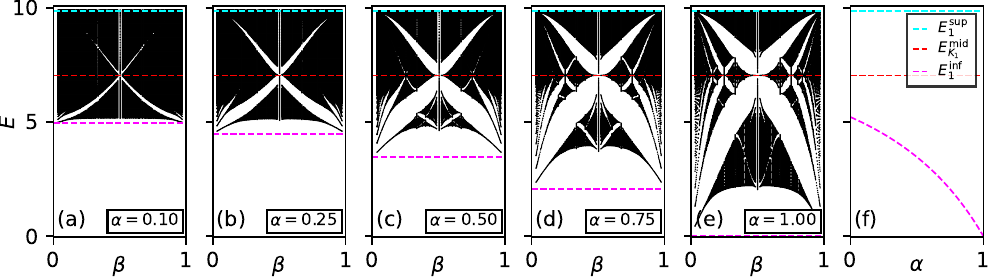}
\caption{\label{fig:fig1Bands} Emergence of the butterfly-like energy spectrum in the lowest projected energy band $E(\beta)$ as the modulation amplitude $\alpha$ is increased. Panels (a)-(e) correspond to $\alpha = 0.1, 0.25, 0.5, 0.75$ and $1$. The energies of the upper bound $E_1^{\mathrm{sup}}$, standing-wave solution $E_{K_1}^{\mathrm{mid}}$ and the lower bound $E_1^{\mathrm{inf}}$ are indicated by cyan, red and magenta dashed lines respectively and their dependence on $\alpha$ is shown in (f). The unmodulated height is fixed at $h_0=10$.}
\end{figure*}

One of the key properties required for topological transport is the presence of isolated bands throughout the entire variable parameter regime, i.e. throughout the range of $\gamma$ and $\Delta$. To determine their existence, controllability and to identify their bounds, we first calculate the energy band structure of the system.

The energy bands of the model are obtained by imposing a periodic boundary condition (PBC) for the coordinate $x$ and solving the eigenvalue problem in quasi-momentum $k$ space (see Appendix \ref{app:PBC}). The system satisfies PBC only if $\beta$ is rational. For numerical calculations we select $\beta = \frac{p}{q}$ with $q\in\{2,3,...,30\}$ and $p$ being all coprimes of $q$ such that $0<\beta<1$. 

We first calculate the set of energy eigenvalues $E(k, \gamma, \beta)$ and look at the band structure as a function of modulation frequency $\beta$, for all values of $k$ and $\gamma$ while $\Delta = 0$. The topological properties of the system will depend only on the presence of band gaps throughout the parameter domain, therefore the relevant information is encoded in the projected bands $E(\beta) = \{E(k, \gamma, \beta) | k\in(-\pi/b, \pi/b],\gamma\in[0,2\pi) \}$ for each $\beta$.
It is important to note that $E(k, \Delta, \beta)$ for $\gamma=0$ shares the same projected band structure $E(\beta) = \{E(k, \Delta, \beta) | k\in(-\pi/b, \pi/b],\Delta\in[0,1) \}$, since varying either $\Delta$ or $\gamma$ periodically covers the whole barrier configuration space albeit in a different manner, leading to same gap structure.
The lowest projected energy band for different modulation strengths $\alpha$ is shown in Fig.~\ref{fig:fig1Bands} (a)-(e). We see that as $\alpha$ increases, the band splits into sub-bands forming a Hofstadter's butterfly-like spectrum \cite{Hofstadter1976Sep}. The higher energy bands behave in a similar fashion (see Appendix \ref{app:PBC}). The number of sub-bands in each band is numerically observed to be $q$. In the even $q$ case, the $q/2$ and $q/2+1$ sub-bands touch at energies $E^{\mathrm{mid}}_{K_n}=K_n^2$ where $K_n$ are the solutions of the equation $\frac{h_0}{2K_n} \sin(K_n)+\cos(K_n) = 0$ with ordering $0 \leq K_1 < K_2 < ... < K_n$. These are the eigenenergies of the system at $k=\pm\frac{\pi}{2}$ when all barrier heights are equal thus they do not depend on $\alpha$ and the wavefunctions form standing waves (see Appendix \ref{app:mappingAA}).

Each of the energy bands, independently of $h_0$, $\alpha$ and $\beta$, is bounded from above by $E^\mathrm{sup}_{n} = (\pi n)^2,\ n\in \mathbb{N}$, where $n$ is the energy band index. This corresponds to the case when the eigenstates of the system effectively do not feel the Dirac-$\delta$ potential since the zeros of the wavefunctions coincide with the positions of the barriers. Thus only the kinetic energy contribution is present.

Finally the lower energy bound $E^\mathrm{inf}_{1}$ of the spectrum is determined by solving the standard Kronig-Penney problem for equal-height barriers of height $h_0(1-\alpha)$ which is the lowest possible energy configuration for a fixed set of parameters (see Appendix \ref{app:mappingAA}). These analytical bounds are depicted with respect to $\alpha$ in Fig.~\ref{fig:fig1Bands} (f).

The symmetry of the energy spectrum with respect to $\beta=0.5$ can be explained by noting that the barriers discretely sample the modulation function $h^{\alpha \beta \gamma \Delta}$. When the modulation frequency is larger than the Nyquist frequency $\beta=0.5$, aliasing occurs since the barriers undersample the modulation signal~\cite{Shannon2006Sep}. This means that frequencies $\beta$ and $1-\beta$ give identical potential modulation up to a sign.

\subsection{\label{subsec:CNumberGamma}Topological properties}

\begin{figure*}
\includegraphics{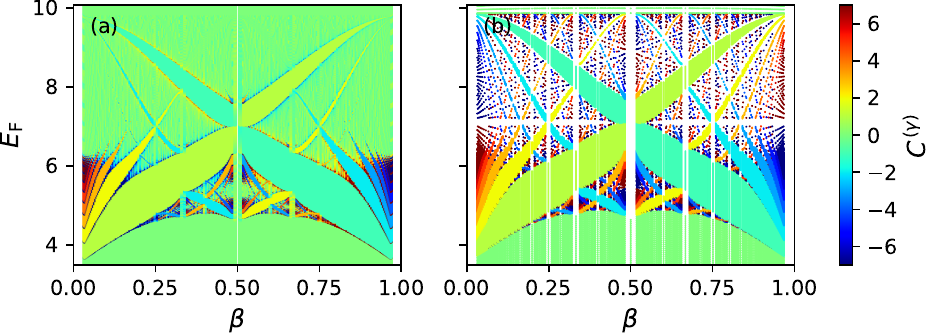}
\caption{\label{fig:fig2Chern} (a) The total Chern number $C^{(\gamma)}$ represented as an in-gap color, indicating the sum of Chern numbers of the sub-bands below Fermi energy $E_\mathrm{F}$. (b) Same color scale is used to mark $t_{n_\mathrm{F}}$ obtained from Diophantine's equation Eq.~(\ref{eq:Diophantine}). The model parameters are $\alpha=0.5$ and $h_0=10$.
}
\end{figure*}

\begin{figure*}
\includegraphics{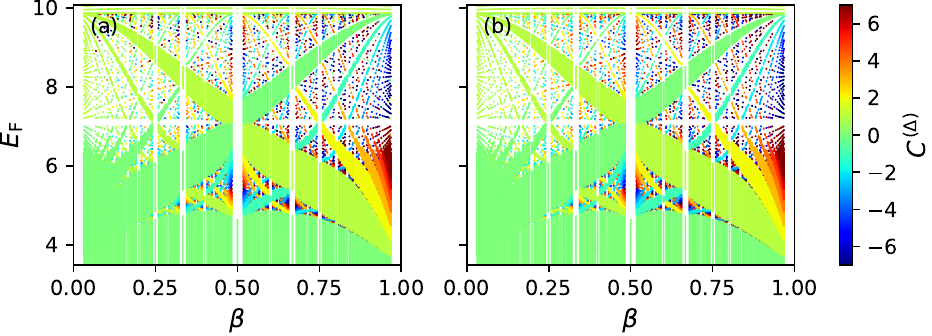}
\caption{\label{fig:fig3Chern} (a) The total Chern number $C^{(\Delta)}$ as an in-gap color, indicating the sum of Chern numbers $\sum_{n=1}^{n_{\mathrm{F}}}C^{(\Delta)}_n$ of the sub-bands below Fermi energy $E_\mathrm{F}$. (b) Coloring given by $s_{n_\mathrm{F}}$ obtained from Diophantine's equation Eq.~(\ref{eq:Diophantine}). The parameters used are $\alpha=0.5$ and $h_0=10$.}
\end{figure*}

The topological properties of the modulated Kronig-Penney system can be captured by treating each modulation parameter as a quasi-momentum of a synthetic dimension.
The real-space quasi-momentum $k$ and a selected modulation parameter then form a closed torus in parameter space, allowing to define the Chern number of isolated energy bands.
This topological invariant provides information about quantized charge transfer in one dimension and can be related to the conductance of a two-dimensional system~\cite{Thouless1982Aug}. 
In this section, we will look at the different transport regimes hosted by the modulated Kronig-Penney model and how they are related to the ones observed in the Harper-Hofstadter model.
The transported charge is obtained by evaluating Chern number in two ways -- either by directly employing the definition which requires information about the states or by using a version of St\v{r}eda formula \cite{Streda1982Aug} which requires only the knowledge of the energy spectrum, thus allowing faster computation.

The first set of periodic parameters that we consider consists of $k$ and $\gamma$. The Chern number associated with the $n$-th sub-band for a selected $\beta$ value is then calculated as~\cite{Fukui2005Jun}:
\begin{equation}
\label{eq:ChernNumber}
    C_n^{(\gamma)} = \frac{1}{2\pi \ri} \int_{\mathrm{BZ}} \! \rd k \! \int_0^{2\pi} \! \rd \gamma \, F_{k\gamma}^{(n)}
\end{equation}
where BZ specifies integration over the first Brillouin zone, $F_{k\gamma}^{(n)}$ is the Berry curvature
\begin{equation}
F_{k\gamma}^{(n)} = \frac{\partial}{\partial k} \left\langle u^{(n)}_{k\gamma} \Bigg| \frac{\partial u^{(n)}_{k\gamma}}{\partial  \gamma} \right\rangle - \frac{\partial}{\partial \gamma} \left\langle u^{(n)}_{k\gamma} \Bigg| \frac{\partial u^{(n)}_{k\gamma}}{\partial  k} \right\rangle
\end{equation}
and $| u^{(n)}_{k\gamma} \rangle$ are the Bloch eigenstates of the sub-band. The second set of parameters that forms a closed torus in the parameter space is $k$ and $\Delta$ with the Chern number calculated as
\begin{equation}
\label{eq:ChernNumber2}
    C_n^{(\Delta)} = \frac{1}{2\pi \ri} \int_{\mathrm{BZ}} \! \rd k \! \int_0^{1} \! \rd \Delta \, F_{k\Delta}^{(n)}.
\end{equation}

The Chern numbers in each case can be evaluated using Eqs.~(\ref{eq:ChernNumber}) and (\ref{eq:ChernNumber2}), however, since the considered model admits a Harper equation representation (see Appendix \ref{app:HH}), a numerically efficient way to obtain the transported charge is to employ St\v{r}eda's formula \cite{Streda1982Aug,Umucalilar2008Feb,Yilmaz2015Jun}.
The transported charge of the completely occupied sub-bands for the $(k,\gamma)$ parameter case are then given by
\begin{equation}
\label{eq:StredaBeta}
    C^{(\gamma)} \equiv \sum_{n=1}^{n_{\mathrm{F}}} C_n^{(\gamma)} = \frac{\partial N(E_{\mathrm{F}})}{\partial \beta},
\end{equation}
where $n_{\mathrm{F}}$ indicates the highest occupied sub-band and $N(E_{\mathrm{F}})$ is the number of states below Fermi energy $E_\mathrm{F}$ divided by total number of states in the band. 
The total charge transfer $C^{(\gamma)}$ of the lowest energy band is calculated numerically using the finite difference method and shown in Fig.~\ref{fig:fig2Chern}(a). The change of $N(E_{\mathrm{F}})$ is divided by the change in modulation frequency of two closest sampled $\beta$ points. 
As long as the given $E_{\mathrm{F}}$ remains in the same energy gap between neighboring $\beta$, the total charge is obtained accurately, however due to the finite sampling of $\beta$ this condition is hard to maintain for small energy gaps.
This is especially prevalent for gaps above $E^{\mathrm{mid}}_{K_1}$ as seen in Fig.~\ref{fig:fig2Chern}(a), where only the charge transfer of the major gaps can be discerned. 
Nevertheless, this approach allows to evaluate the Chern numbers for a large range of $\beta$ quickly. 
In contrast, the transferred charge $C^{(\Delta)} \equiv \sum_{n=1}^{n_{\mathrm{F}}} C_n^{(\Delta)}$ in $(k,\Delta)$ space is obtained using Eq.~(\ref{eq:ChernNumber2}) and is shown in Fig.~\ref{fig:fig3Chern}(a). Integrating the Berry curvature provides more accurate results since the computation is performed for a fixed $\beta$, however, numerical errors can still appear when the gap between the energy bands is small and finer parameter discretization is needed.

The two charge transfer regimes shown in Figs.~\ref{fig:fig2Chern}(a) and \ref{fig:fig3Chern}(a) share the same butterfly-like projected bandwidth structure although their topological properties are different. 
In particular, the regimes are connected through the Diophantine equation (see Appendix \ref{app:HH})
\begin{equation}
    pC^{(\gamma)} +qC^{(\Delta)} = n_\mathrm{F},
\end{equation}
for $n_\mathrm{F}$ filled bands.
This allows to establish a connection between the one-dimensional Kronig-Penney model under parameter variation and the charge transfer picture of the two-dimensional lattice with electrons in a magnetic field. 
Such a system is known as the Hofstadter model and the information about its transport is encoded in the Diophantine equation of the form
\begin{equation}
\label{eq:Diophantine}
    p t_{n_\mathrm{F}} + q s_{n_\mathrm{F}} = n_\mathrm{F},
\end{equation}
with conditions $2|t_{n_\mathrm{F}}| < q$ and $s_{n_\mathrm{F}} \in \mathbb{Z}$. It uniquely determines the Hall conductance $\sigma_{\mathrm{H}}^{(n_{\mathrm{F}})}=\frac{e^2}{2\pi\hbar}t_{n_\mathrm{F}}$ of a square lattice pierced by a dimensionless rational magnetic flux $\phi_{\mathrm{flux}} = \frac{p}{q}$ for the system occupied by particles up to the $n_\mathrm{F}$-th energy band \cite{Thouless1982Aug,Wannier1978Aug,MacDonald1983Dec,Dana1985Aug}. The integer $s_{n_\mathrm{F}}$ is the gap label that fixes the flux-independent part of the integrated density of states $N(E_\mathrm{F})$ once $t_{n_\mathrm{F}}$ is determined.
It can be interpreted as a Chern number associated with the charge transfer under the translation of the periodic potential \cite{Kohmoto1992Aug}.
We map the calculated charge $t_{n_\mathrm{F}}$ of the Hofstadter model to the energy gaps of the modulated Kronig-Penney model assuming that $\beta \equiv \phi_\mathrm{flux}$. The resulting gap-coloring based on the total Chern number $t_{n_\mathrm{F}}$ corresponding to the Hall conductance of the lowest energy band is shown in Fig.~\ref{fig:fig2Chern}(b). Comparing it with the charge transfer obtained from St\v{r}eda's formula at the well-resolved gaps (Fig.~\ref{fig:fig2Chern}(a)) we see that the predicted Chern numbers of both models are in good agreement. Comparing $C^{(\Delta)}$ of Fig.~\ref{fig:fig3Chern}(a) to $s_{n_\mathrm{F}}$ obtained from Eq.~(\ref{eq:Diophantine}) and mapped to the gaps of the modulated model (Fig.~\ref{fig:fig3Chern}(b)) we get matching Chern numbers up to numerical precision as well. 

The origin of the topological similarity between the modulated sub-wavelength barrier Hamiltonian and the Harper-Hofstadter model lies in the underlying difference equation structure that both models share. The 2D Harper-Hofstadter lattice model can be mapped to a 1D Harper equation \cite{Harper1955Oct} (also known as the Aubry-Andr\'{e} model \cite{Aubry1980}) by a proper gauge choice \cite{Kraus2012Sep2} while the modulated Kronig-Penney model reduces to a modified Harper equation by employing the Bethe ansatz (see Appendix \ref{app:mappingAA} and \ref{app:HH}). Even though the dispersion relations between the models differ, the underlying topological properties are retained as long as the energy gaps do not close. 

Two aspects of the modulated Kronig-Penney model go beyond the standard Harper-Hofstadter setting.
First, as a continuum system rather than a tight-binding lattice, its spectrum is not bounded from above, providing access to topological structure in higher-energy bands.
Second, the model admits an extended modulation parameter space: whereas in the Hofstadter problem the gap-labeling integer $s_{n_{\mathrm F}}$ is fixed once the band Hall response is specified, here the associated invariant $C^{(\Delta)}$ is directly accessible experimentally through topological pumps by changing parameter $\Delta$, as discussed below. 
Together, these features enable transport responses that differ qualitatively from the canonical lattice case, e.g. filling the entire sub-bands of the lowest band one can obtain a nontrivial net charge $C^{(\Delta)}=1$, in contrast to the usual Hofstadter model where filling all the bands yields $C^{(\gamma)}=0$.

\subsection{\label{subsec:realSpaceTransport}Adiabatic transport in real-space}

\begin{figure}
\includegraphics{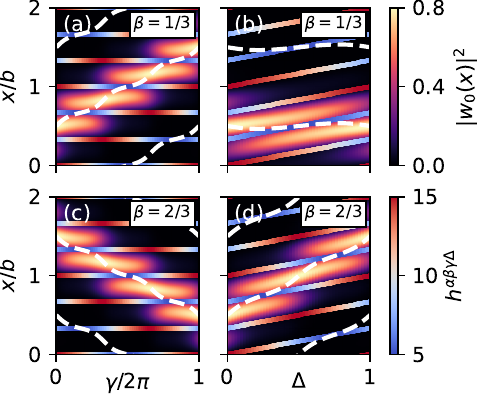}
\caption{\label{fig:fig4Wannier} Spatial transfer of localized Wannier function density $|w_0(x)|^2$ under adiabatic change of $\gamma$ and $\Delta$ when the lowest energy sub-band is filled for modulation $\beta = \frac{1}{3}$ in (a,b) and $\beta=\frac{2}{3}$ in (c,d). Two elementary cells are shown with barrier positions and heights indicated by $h^{\alpha\beta\gamma\Delta}$. The dashed white curves are position expectation values of Wannier functions of neighboring elementary cells.  }
\end{figure}

The resulting non-trivial topology of the model leads to quantized density charge transfer in real-space under the adiabatic change of $\gamma$ and $\Delta$. This is known as Thouless pumping which has been observed in a variety of cold atom and condensed matter systems \cite{Thouless1983May,Nakajima2016Apr,Citro2023Feb}. 

To illustrate transport in our system we calculate the position expectation value of localized Wannier functions \cite{Marzari2012Oct} of the occupied bands during a single periodic evolution of the parameters, i.e. a single pumping cycle. For systems with PBC, the position operator $\hat{x}$ is not compatible with translational symmetry on a ring, making its expectation value ill-defined and origin-dependent \cite{Resta1998Mar}. A suitable expression that respects the translational symmetry can be constructed using the unitary operator $\re^{\ri\frac{2\pi}{L}\hat{x}}$, defining the position up to modulo of the system length \cite{Asboth2016}
\begin{equation}
    \langle x \rangle = \frac{L}{2\pi}\operatorname{Im}\left[\log\left(\left\langle \re^{\ri\frac{2\pi}{L}\hat{x}}\right\rangle \right)\right].
\end{equation}
Here $L$ is the length of a ring that consists of $M_{\mathrm{cell}}$ elementary cells of length $b$. Evaluating this expression for the Wannier functions gives the Wannier center position.

We select two modulation frequencies $\beta=\frac{1}{3}$ and $\beta=\frac{2}{3}$, which are numerically feasible and capture distinct transport regimes during the variation of $\gamma$ and $\Delta$, including transport in opposite directions as well as the absence of net transport. 
We focus on the lowest sub-band of the lowest energy band and assume it is completely filled. 
The calculated Chern numbers for the considered cases are shown in Table \ref{tab:betaChern}.
\begin{table}[tb]
\centering
\caption{Chern numbers of the first energy band for the parameter spaces $(k,\gamma)$ and $(k,\Delta)$ at selected $\beta$. The labels (a-d) indicate the matching transport cases of Fig.~\ref{fig:fig4Wannier}.}
\label{tab:betaChern}
\begin{tabular}{c|c|c|}
\cline{2-3}
                       & $C_1^{(\gamma)}$ & $C_1^{(\Delta)}$ \\ \hline
\multicolumn{1}{|l|}{$\beta = 1/3$} &  \quad $1$ (a) & \quad $0$  (b)  \\ \hline
\multicolumn{1}{|l|}{$\beta = 2/3$} & \ $-1$ (c) & \quad $1$ (d) \\ \hline
\end{tabular}
\end{table}
The accompanying Fig.~\ref{fig:fig4Wannier} illustrates the spatial transport of localized densities for these regimes. In Fig.~\ref{fig:fig4Wannier} (a) and (c), the barriers remain in fixed positions and the density of Wannier functions $w_0(x)$ is transported across the barriers with lowest heights while varying $\gamma$. The transport in different directions for $\beta=\frac{1}{3}$ and $\beta=\frac{2}{3}$ is observed because the modulation envelope is offset by $\frac{2\pi}{3}$ in (a) and $-\frac{2\pi}{3}$ in (b) for each subsequent barrier, leading to the movement of the modulation minima to either right or left during the pump cycle. In Fig.~\ref{fig:fig4Wannier} (b) and (d), the barriers are shifted by $\Delta$ in space as well.
The Wannier centers (dashed white curves) are moved through the barriers with lowest heights resulting in no transport between elementary cells in (b) and transport to the neighboring elementary cell in (d).  In all of the cases, the density of a selected Wannier function, after completing a single pumping cycle, gets translated by an integer number of elementary cells. This integer corresponds to the Chern number with the sign indicating the transport direction as expected (see Table \ref{tab:betaChern}).

In ultracold atom systems, the analyzed transport can be observed using standard techniques such as measuring the center of mass movement of an atom cloud under pumping~\cite{Nakajima2016Apr}. The Chern number can be also obtained from density-profile measurements by monitoring the bulk particle density under the change of modulation frequency~\cite{Repellin2020Dec,Leonard2023Jul}. 

\subsection{Experimental realization}

\begin{figure}[ht!]
\includegraphics[width=0.98\columnwidth]{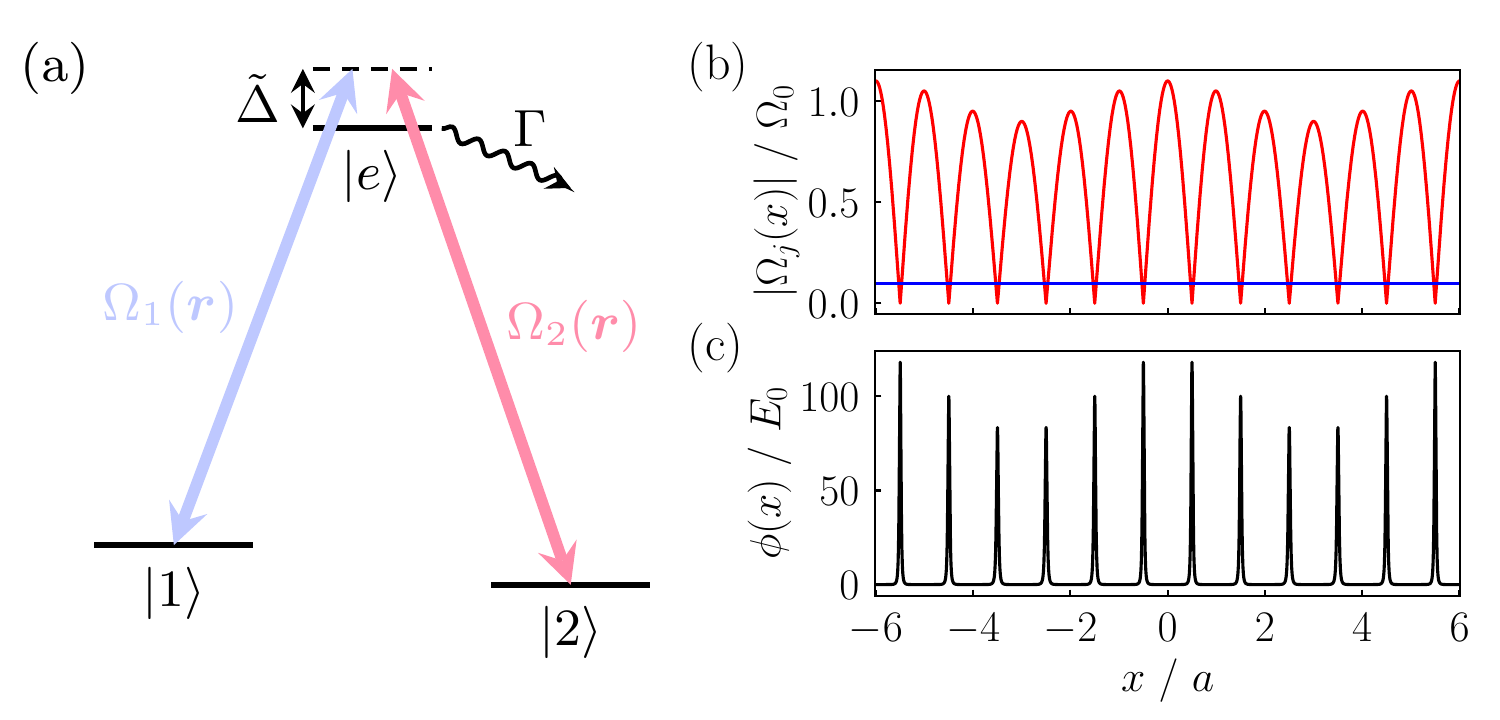}
\caption{\label{fig:figLambda} (a) Lambda atom-light coupling configuration for ultracold atoms. 
(b) Spatial dependence of the  modulus of Rabi frequencies $\Omega_{1}$ and $\Omega_{2}$ obtained from Eqs.~\eqref{eq:Omega1-1dsubwave}-\eqref{eq:hx-generalized} for $\epsilon=0.1$, $\alpha=0.1$, $\beta=\frac{1}{3}$, $\gamma=0$, $k=1$, $\theta=0$. (c) Position dependence of geometric scalar potential $\phi$ for aforementioned $\Omega_j$ configuration, obtained from Eq.~\eqref{eq:phi-dark} for the same parameters.}
\end{figure}

Let us also propose a concrete experimental scheme to realize the aforementioned Kronig-Penney-type model. For this we consider ultracold atoms in a $\Lambda$ configuration with two ground states $|1\rangle$ and $|2\rangle$ and an excited state $|e\rangle$, as depicted in Fig.~\ref{fig:figLambda}(a). In general, the excited state experiences spontaneous emission, characterized by the decay rate $\Gamma$. The low-lying states $|1\rangle$ and $|2\rangle$ are off-resonantly coupled (with detuning $\tilde\Delta$) to the excited state $|e\rangle$ by laser fields, which are characterized by spatially dependent Rabi frequencies $\Omega_1(\mathbf{r})$ and $\Omega_2(\mathbf{r})$ respectively. In what follows, we will usually keep the position dependence implicit.

Applying the rotating wave approximation \cite{Scully2008}, the standard atom-light coupling Hamiltonian can be written as
\begin{equation}
\hat{V}/\hbar=\sum_{j=1}^{2}\left(\frac{\Omega_{j}}{2}\left|e\right\rangle \left\langle j\right|+{\rm H.c.}\right)-\left(\tilde\Delta+\frac{\ri}{2}\Gamma\right)\left|e\right\rangle \left\langle e\right|\,,\label{eq:V-definition}
\end{equation}
which supports a dark state solution with eigenvalue $0$ which is given by~\cite{Dalibard11RMP, MorrisShore1983PRA}
\begin{equation}
\left|D\right\rangle =\frac{1}{\Omega}\left(\Omega_{2}\left|1\right\rangle -\Omega_{1}\left|2\right\rangle \right)\,,
\end{equation}
where $\Omega=\sqrt{\left|\Omega_{1}\right|^{2}+\left|\Omega_{2}\right|^{2}}\label{eq:Omega}$.

When the amplitude of the Rabi frequencies is much larger than the characteristic kinetic energy, the atoms adiabatically follow the dark state, i.e., $|\psi(\mathbf{r})\rangle \approx \psi_D(\mathbf{r})\,|D(\mathbf{r})\rangle$ \cite{Dalibard11RMP, Juz05PRA, Goldman2014, Juz05JPB}. Then, the dark state wavefunction $\psi_D$ is governed by the dark state Hamiltonian
\begin{equation}
\hat{H}_{{\rm D}}=-\frac{\mathrm{d}^2}{\mathrm{d}x^2}+\phi\,,\label{eq:H_D}
\end{equation}
where $\phi$ is the geometric scalar potential \cite{Juz05PRA, Juz05JPB}
\begin{equation}
\phi=\frac{\mathbf{\nabla}\zeta^{*}\cdot\mathbf{\nabla}\zeta}{\left(1+|\zeta|^{2}\right)^{2}}\,,\label{eq:phi-dark}
\end{equation}
and $\zeta=\Omega_{1}/\Omega_{2}$ is the Rabi frequency ratio. Note that for a one-dimensional system, the vector potential can always be gauged away and thus, it is not shown. Henceforth, we will take the Rabi frequencies to be purely real.

Looking at Eq.~\eqref{eq:phi-dark}, we see that the amplitude of the scalar potential depends on the spatial variation of the Rabi frequency ratio. Consequently, a sub-wavelength Kronig-Penney lattice can be engineered if one of the Rabi frequencies periodically goes to zero, while the other has a much smaller constant amplitude~\cite{Gvozdiovas2021Dec, Lacki2016Nov}. To additionally obtain modulated Dirac-$\delta$ scatterer heights, one may add an amplitude envelope on one (or both) of the Rabi frequencies.

More concretely, consider the following class of configurations
\begin{align}
\Omega_1\left(x\right) &= \Omega_0 f_1\left(x\right) \sin\left(kx+\theta\right)\,,
\label{eq:Omega1-1dsubwave}\\
\Omega_2\left(x\right) &= \epsilon\Omega_0 f_2\left(x\right)\,,
\end{align}
where $\Omega_0$ is the Rabi frequency amplitude, $\epsilon$ is the amplitude ratio, $k$ is the wavevector, $\theta$ is the spatial phase, and $f_i\left(x\right)$ are functions of order unity with no real roots ($i\in\{1,2\}$).

Then consider the vicinity of a zero of $\Omega_1\left(x\right)$ at $x=x_j$. One can show that the geometric scalar potential approaches a Dirac-$\delta$ peak at $x=x_j$ with amplitude $\pi\delta\left(x-x_j\right)/2\epsilon_j$ as $\epsilon\rightarrow0$, where $\epsilon_j = \epsilon f_2\left(x_j\right)/f_1\left(x_j\right)$ \cite{Lacki2023Nov}. Since $\Omega_1\left(x\right)$ is periodic, one obtains a sub-wavelength barrier array with spatial period $a=\pi/k$.

To obtain the considered Kronig-Penney lattice, given by Eq.~\eqref{eq:model}, one can take the following envelope functions
\begin{equation}
f_1\left(x\right)=h^{\alpha \beta \gamma \Delta}\left(x\right) / h_0\,,\quad f_2\left(x\right)=1\,, 
\end{equation}
where
\begin{equation}
h^{\alpha \beta \gamma \Delta}\left(x\right) = h_0 [1+\alpha\cos(2\pi\beta x^{\Delta}-\gamma)]\,,
\label{eq:hx-generalized}
\end{equation}
and $x^{\Delta}=x+\Delta$. Of course, one has $h^{\alpha \beta \gamma \Delta}\left(x_j\right)=h^{\alpha \beta \gamma \Delta}_j$.

For clarity, the Rabi frequencies and  the resulting geometric scalar potential are shown in Fig.~\ref{fig:figLambda}(b) and (c) respectively. Note that in principle, there are many possible configurations to achieve the desired lattice since the Dirac-$\delta$ amplitude is determined only by the envelope function ratio at $x=x_j$ and not the individual envelope function values.

\section{\label{sec:Summary}Summary and outlook}

In this work we have demonstrated that a spatially modulated sub-wavelength barrier lattice exhibits nontrivial band topology characteristic of Hofstadter-type systems. The modulation of the barrier strengths fragments the bands into multiple sub-bands and produces a Hofstadter-butterfly-like spectrum with analytically controllable bounds. Treating the modulation parameters as synthetic dimensions, we identified two pumping geometries, $(k,\gamma)$ and $(k,\Delta)$, which display distinct quantized transport regimes under adiabatic variation. The associated Chern numbers, obtained from the Berry curvature and a St\v{r}eda-type relation, are linked by a Harper-Hofstadter-like Diophantine equation, providing a connection between the modulated continuum model and the quantum Hall lattice problem. These bulk invariants are directly reflected in real-space Thouless pumping, where Wannier centers shift by an integer number of elementary cells over a pumping cycle.

We have also outlined a concrete implementation based on dark-state optical potentials in a three-level $\Lambda$ configuration, where position-dependent Rabi frequencies generate an array of modulated Dirac-$\delta$ barriers. Such setups are compatible with current ultracold atomic gas experiments, making modulated Kronig-Penney systems a practical platform for realizing Hofstadter-type topology in reduced dimensionality. While our analysis focused on bulk properties, the flexibility in engineering barrier configurations and band topologies suggests straightforward extensions to topological interfaces and edge-state physics. In future work, it will be of particular interest to investigate how interactions and disorder modify the topological transport identified here, and to develop controlled non-adiabatic driving protocols for manipulating transport beyond the strictly adiabatic regime. 

\begin{acknowledgments}
This work was supported by the Okinawa Institute of Science and Technology Graduate University (OIST) and utilized the computing resources of the Scientific Computing and Data Analysis section of Core Facilities at OIST. This work was initiated by SHINKA grant, a joint grant between OIST and Tohoku University, and supported by the JSPS Bilateral Program No.~JPJSBP120244202. DB is supported by the Research Council
of Lithuania (RCL) Grant No. S-LJB-24-2 and the JSPS Bilateral
Program No. JPJSBP120244202. This work has also
received funding from COST Action POLYTOPO CA23134,
supported by COST (European Cooperation in Science and
Technology). TO is supported by JSPS KAKENHI Grant No. JP24K00548 and JST PRESTO Grant No. JPMJPR2353.
\end{acknowledgments}

\appendix

\section{\label{app:PBC}Hamiltonian matrix elements of periodic boundary modulated Kronig-Penney model}

Let the height modulation frequency in Eq.~(\ref{eq:model}) be rational, i.e $\beta = \frac{p}{q}$ where $p$ and $q$ are coprime integers. In this case, for a given $\beta$, an elementary supercell of length $b$ exists. To solve the eigenvalue problem we use Bloch's theorem stating that the eigenfunctions of a periodic system have the form $\psi^{(n)}_k(x) = \re^{\ri kx}u^{(n)}_{k}(x)$. Here $n$ is the quantum number labeling the energy bands, $u^{(n)}_{k}(x) = u^{(n)}_{k}(x + b)$ is the periodic part of the Bloch wavefunction and $k$ is the quasi-momentum. If periodic boundary conditions are imposed quasi-momentum takes discrete values $k=\frac{2\pi \ell }{M_{\mathrm{cell}} b}$ with $\ell \in \{-\frac{M_{\mathrm{cell}}}{2},\frac{M_{\mathrm{cell}}}{2} + 1, ..., \frac{M_{\mathrm{cell}}}{2}-1 \}$ and $M_{\mathrm{cell}}$ being the number of supercells. In the infinite lattice case $k$ is continuous. Acting with the model Hamiltonian on the wavefunction ansatz we get a set of decoupled equations for each $k$ with the periodic part of the wavefunction as the eigensolution
\begin{equation}
   \left[\left(\ri\frac{\rd}{\rd x} - k \right)^2 + V(x)\right] u^{(n)}_{k}(x) = E_n u^{(n)}_{k}(x), 
\end{equation}
with $V(x) = \sum_{j\in \mathcal{M}} h^{\alpha \beta \gamma \Delta}_j  \delta(x-x_j^{\Delta}) $ and energy eigenvalues $E_n$. We can write the matrix representation of this Hamiltonian in plane-wave basis $\phi_m(x) = \frac{1}{\sqrt{b}}\re^{\ri \frac{2\pi m}{b}x}$, $m \in \mathbb{Z}$. The solutions are then $u^{(n)}_{k}(x) = \sum_{m=-\infty}^{\infty} c_{mk}^{(n)} \phi_m(x)$, with coefficients $c_{mk}^{(n)}\in \mathbb{C}$ obtained by diagonalizing the Hamiltonian matrix. The explicit form of the matrix elements is
\begin{align}
    &H_{ml} \equiv \int_0^b\! \rd x \, \phi_m^*(x) H \phi_l(x) = \left(k + \frac{2\pi l}{b} \right)^2 \delta_{ml} \nonumber \\
    &+\frac{1}{b} \sum_{j\in \mathcal{M}_\mathrm{cell}} h^{\alpha \beta \gamma \Delta}_j \re^{\ri \frac{2\pi (l-m)}{b} x_j^{\Delta}},
\end{align}
where the sum is over the set $\mathcal{M}_\mathrm{cell}$ of barriers indexed in a single supercell.

\begin{figure}
\includegraphics{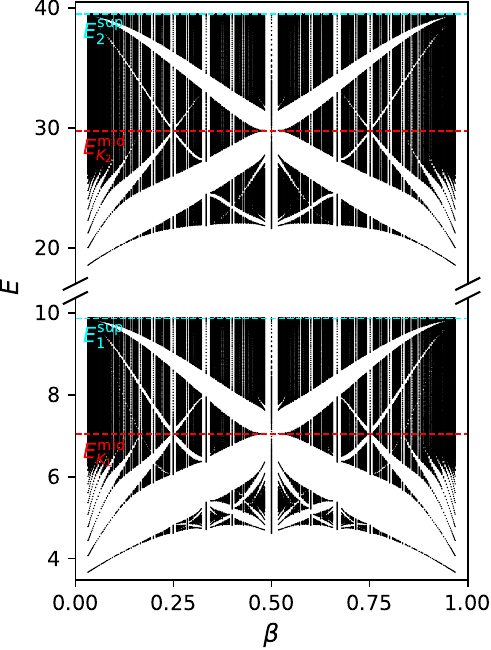}
\caption{\label{fig:fig1BandsApp} Splitting of the two lowest energy bands $E(\beta)$ under the change of modulation frequency for $h_0=10$, $\alpha=0.5$, $\Delta = 0$. The dependence on $k$ and $\gamma$ is projected onto the energy axis for each $\beta$. The number of elementary cells is taken to be $\mathcal{M}_\mathrm{cell} =50$. Cyan dashed line indicates the upper bound of the energy $E_j^{\mathrm{sup}}$ and red dashed line indicates the standing-wave solution energy $E_{K_j}^{\mathrm{mid}}$ for each band $j\in\{1,2\}$.}
\end{figure}

For numerical calculations, the plane-wave number cutoff is $|m_{\mathrm{cutoff}}| = 1000$ modes. The dispersions $E(k,\gamma)$ and $E(k,\Delta)$, for fixed $\alpha$ and $\beta$, are computed by discretizing the parameter ranges $k\in [-\frac{\pi}{b},\frac{\pi}{b})$, $\gamma \in [0,2\pi)$ and $\Delta \in [0,1)$ into $50$ evenly spaced points. The bandwidths of both dispersion spectra coincide and the energy band projections for each $\beta$ are shown for the two lowest bands in Fig.~\ref{fig:fig1BandsApp}. The second band retains the features of the first band but the fractalization of the gaps is less pronounced. 

\section{\label{app:mappingAA} Determining analytical limits of the modulated Kronig-Penney model}

We start with the Hamiltonian (\ref{eq:model}) of the main text
\begin{equation}
\label{eq:modelApp}
       H = -\frac{\rd^2}{\rd x^2} +   \sum_{j\in \mathcal{M}} h_j   \delta(x-x_j), 
\end{equation}
where we have dropped the indices indicating parameter dependence for brevity. We assume that the wavefunction between the barriers supports a piece-wise plane-wave solution, i.e.
\begin{equation}
    \psi(x) = A_j \re^{\ri \sqrt{E} x} + B_j \re^{-\ri \sqrt{E} x}, \ \mathrm{for} \  x_{j-1} < x \leq x_j, \  j\in \mathcal{M},
\end{equation}
given the eigenproblem $H\psi(x) = E\psi(x)$. At the position of the $j$-th barrier we use the continuity of the wavefunction
\begin{equation}
    \lim_{\varepsilon \rightarrow0} \left[\psi(x_j+\varepsilon) - \psi(x_j-\varepsilon) \right] = 0,
\end{equation}
which leads to the following condition for the first derivative
\begin{equation}
    \lim_{\varepsilon \rightarrow0} \left[\frac{\rd \psi(x)}{\rd x}\Bigg|_{x_j+\varepsilon}  - \frac{\rd \psi(x)}{\rd x}\Bigg|_{x_j-\varepsilon} \right] = h_j \psi(x_j).
\end{equation}
Using these relations one arrives at the relations of coefficients between neighboring regions
\begin{equation}
\begin{aligned}
\tilde{A}_{j+1} \re^{-\ri \sqrt{E} a} &= \left(1-\frac{\ri h_j}{2\sqrt{E}} \right)\tilde{A}_{j}- \frac{\ri h_j}{2\sqrt{E}}\tilde{B}_{j}, \\
\tilde{B}_{j+1} \re^{\ri \sqrt{E} a} &= \frac{\ri h_j}{2\sqrt{E}}  \tilde{A}_{j}+ \left(1+\frac{\ri h_j}{2\sqrt{E}} \right)\tilde{B}_{j}.
\end{aligned}
\label{eq:tMat}
\end{equation}
Here $\tilde{A}_{j} \equiv A_j \re^{\ri \sqrt{E}x_j}$, $\tilde{B}_{j} \equiv B_j \re^{-\ri \sqrt{E}x_j}$ and $a$ is the separation between nearest barriers. Noting that $\psi_j \equiv\psi(x_j) = \tilde{A}_{j} + \tilde{B}_{j}$, the Eqs.~(\ref{eq:tMat}) can be expressed as
\begin{equation}
\label{eq:modulatedKPTM}
    \frac{1}{2}\left(\psi_{j+1} + \psi_{j-1}\right) - \frac{h_j \sin(\sqrt{E}a)}{2\sqrt{E}}\psi_j = \cos(\sqrt{E}a)\psi_j.
\end{equation}

A feature of the butterfly-like spectrum can be immediately inferred where the barriers are equal for a given $\beta$ during parameter change. This corresponds to the formation of standing waves with zero group velocity, i.e., $\frac{1}{2}\left(\psi_{j+1} + \psi_{j-1}\right)=0$, which leads to the condition $\frac{h \sin(\sqrt{E_n}a)}{2\sqrt{E_n}} + \cos(\sqrt{E_n}a) = 0$ for the $n$-th band, when all barrier heights are equal to $h$. The energy solutions correspond to $E_n \propto E^{\mathrm{mid}}_{K_n}$ in the main text. One can also recover the standard Kronig-Penney result by Fourier expanding the coefficients $\psi_j = \sum_k \re^{\ri k x_j} \psi_{k}$, leading to 
\begin{equation}
    \cos(ka) - \frac{h \sin(\sqrt{E}a)}{2\sqrt{E}} = \cos(\sqrt{E}a).
\end{equation}
Setting the quasi-momentum value $k=0$ and solving the equation allows to find the lower bound $E^\mathrm{inf}_{1}$ of the energy spectrum for the butterfly-like structure. 

\section{\label{app:HH} Relation to Harper-Hofstadter equation}

The modulated Kronig-Penney Eq.~(\ref{eq:modulatedKPTM}) derived in the previous section can be rewritten as
\begin{equation}
    \label{eq:HarperForm}
    \psi_{j+1} + \psi_{j-1}   + g_1(E)\cos(2\pi\beta x_j - \varphi) \psi_{j} = g_2(E) \psi_{j},
\end{equation}
where we have denoted the modulation phase as
\begin{equation}
    \varphi = \gamma - 2\pi\beta\Delta,
\end{equation}
and the energy dependent terms by
\begin{align}
    g_1(E)  &\equiv - \frac{ \alpha h_0 \sin(\sqrt{E}a) }{\sqrt{E}},  \\
    g_2(E) &\equiv 2\cos(\sqrt{E} a) + \frac{ h_0 \sin(\sqrt{E}a) }{\sqrt{E}}.
\end{align}
The obtained Eq.~(\ref{eq:HarperForm}) shares the discrete difference equation structure of the Harper-Hofstadter model with an energy dependent modulation coefficient. The modulation frequency $\beta$ plays an analogous role to the dimensionless magnetic flux $\phi_\mathrm{flux}$ in the 2D Hofstadter model written in the Harper equation form \cite{Kohmoto1992Aug}. A Thouless pump cycle in the Harper equation is realized by adiabatically changing the phase parameter $\varphi$ from $0$ to $2\pi$. The charge transported during such a period for a selected $n$ band is equal to the Chern number
\begin{equation}
    C_n^{(\varphi)} = \frac{1}{2\pi \ri} \int_{\mathrm{BZ}} \! \rd k \! \int_0^{2\pi} \! \rd \varphi \, F_{k\varphi}^{(n)},
\end{equation}
where the superscript indicates the pumping parameter considered. We immediately see that when $\Delta$ is fixed and the pumping is induced by changing $\gamma$, the pumped charge precisely coincides with the Hofstadter model case, i.e. $C_n^{(\gamma)} = C_n^{(\varphi)}$. It is known that for $n_F$ occupied energy sub-bands of the Hofstadter model, the transported charge can be related by St\v{r}eda-Widom formula
\begin{equation}
    \frac{e^2}{2\pi \hbar}\sum_{n=1}^{n_\mathrm{F}}C_n^{(\varphi)} =  e   \frac{\partial \rho (E_{\mathrm{F}})}{\partial B} ,
\end{equation}
where $e$ is the electron charge, $B$ -- magnetic field and $\rho (E_{\mathrm{F}}) = \frac{N(E_{\mathrm{F}})}{A_\mathrm{cell}}$ is the particle density with $A_\mathrm{cell}$ -- elementary cell area and $N(E_{\mathrm{F}})$ -- density of occupied states per band, given that all energy levels below $E_{\mathrm{F}}$ are filled. Expressing the equation in terms of dimensionless magnetic flux $\phi_\mathrm{flux} = \frac{cBA_\mathrm{cell}}{2\pi \hbar}$, we get
\begin{equation}
    C^{(\varphi)}\equiv \sum_{n=1}^{n_\mathrm{F}}C_n^{(\varphi)} = \frac{\partial N(E_{\mathrm{F}})}{\partial \phi_\mathrm{flux}}.
\end{equation}
Analogous relation is valid for the Kronig-Penney model if the dimensionless flux is replaced by the spatial modulation frequency of the barriers, provided that the functions $g_1(E)$ and $g_2(E)$ only smoothly deform the energy bands without closing the gap
\begin{equation}
    C^{(\gamma)} = \frac{\partial N(E_{\mathrm{F}})}{\partial \beta}.  
\end{equation}
This is the Eq.~(\ref{eq:StredaBeta}) used in the main text.

If $\gamma$ is fixed and $\Delta$ is varied, the Chern number for a single pump cycle is
\begin{equation}
    C_n^{(\Delta)} = \frac{1}{2\pi \ri} \int_{\mathrm{BZ}} \! \rd k \! \int_0^{1} \! \rd \Delta \, F_{k\Delta}^{(n)}.
\end{equation}
For rational modulation $\beta = p/q$, the transported charge $C^{(\Delta)}$ can be related to $C^{(\gamma)}$ by noting that performing the $\Delta$ pump cycle $q$ times shifts the phase $\varphi$ by $-2\pi p$ and the barriers are shifted by an elementary lattice cell. Shifting the barriers transports $n_\mathrm{F}$ charges corresponding to the filled number of sub-bands~\cite{Thouless1983May} and the phase change contributes $-pC^{(\gamma)}$, leading to the total charge transport 
\begin{equation}
    qC^{(\Delta)} = n_\mathrm{F} - pC^{(\gamma)}.
\end{equation}
Rearranging gives the Diophantine equation for the considered modulated Kronig-Penney model
\begin{equation}
    pC^{(\gamma)} + qC^{(\Delta)} = n_\mathrm{F},
\end{equation}
or alternatively
\begin{equation}
    C^{(\Delta)} + \beta C^{(\gamma)} = N(E_{\mathrm{F}}),
\end{equation}
which can also be obtained by integrating St\v{r}eda's Eq.~(\ref{eq:StredaBeta}).

\providecommand{\noopsort}[1]{}\providecommand{\singleletter}[1]{#1}%

\end{document}